\documentclass[conference,10pt]{IEEEtran}
\usepackage{epsfig,rotating,setspace,latexsym,amsmath,epsf,amssymb,amsfonts,bm,theorem,subfigure,epstopdf}
\usepackage{cite,authblk}
\usepackage{bbm}
\usepackage{algorithm}
\usepackage[noend]{algpseudocode}

\algrenewcommand\algorithmicforall{\textbf{foreach}}
\algrenewcommand\algorithmicindent{.8em}
\usepackage{color}
\usepackage{mathtools}

\newtheorem{theorem}{Theorem}

\newtheorem{corollary}{Corollary}
\newenvironment{Proof}[1]{\medskip\par\noindent{\bf Proof:\,}\,#1}{{\mbox{\,$\blacksquare$}\par}}
\IEEEoverridecommandlockouts
\allowdisplaybreaks

\begin{document}

	\title{Timely Communication in Federated Learning \thanks{This work was supported by NSF Grants CCF 17-13977 and ECCS 18-07348.}}
	
	\author{Baturalp Buyukates \qquad Sennur Ulukus\\
	\normalsize Department of Electrical and Computer Engineering\\
	\normalsize University of Maryland, College Park, MD 20742\\
	\normalsize  \emph{baturalp@umd.edu} \qquad \emph{ulukus@umd.edu}}

	\maketitle

\begin{abstract}
We consider a federated learning framework in which a parameter server (PS) trains a global model by using $n$ clients without actually storing the client data centrally at a cloud server. Focusing on a setting where the client datasets are fast changing and highly temporal in nature, we investigate the timeliness of model updates and propose a novel timely communication scheme. Under the proposed scheme, at each iteration, the PS waits for $m$ available clients and sends them the current model. Then, the PS uses the local updates of the earliest $k$ out of $m$ clients to update the global model at each iteration. We find the average age of information experienced by each client and numerically characterize the age-optimal $m$ and $k$ values for a given $n$. Our results indicate that, in addition to ensuring timeliness, the proposed communication scheme results in significantly smaller average iteration times compared to random client selection without hurting the convergence of the global learning task.
\end{abstract}
 
\section{Introduction}
\begin{figure}[t]
	\centering  \includegraphics[width=0.93\columnwidth]{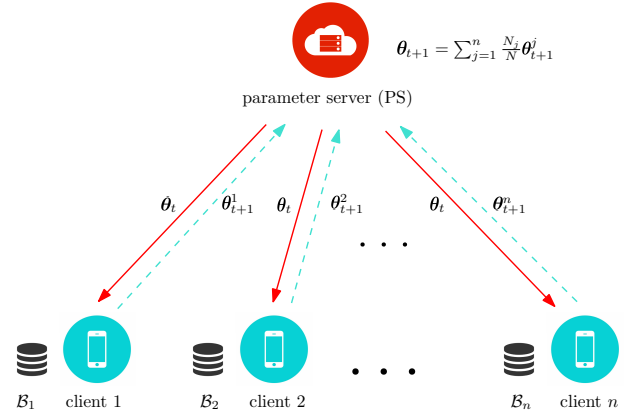}
	\caption{Federated learning model where a parameter server (PS) trains a learning model using $n$ clients without actually storing clients' data centrally. }
	\label{fig:FL_update}
	\vspace{-0.5cm}
\end{figure}

Introduced in \cite{McMahan17}, federated learning (FL) is a distributed learning framework, where a parameter server (PS) iteratively trains a global model using rich client (user) datasets that are privacy-sensitive and large in quantity without actually storing them centrally at the data center. At each iteration, the PS distributes the current model to the clients. Each client performs the learning task locally using its own dataset and sends its model update to the PS, which then aggregates the results and updates the model (see Fig.~\ref{fig:FL_update}). Some promising applications of FL are image classification and next-word prediction \cite{McMahan17}, human mobility prediction \cite{Feng20}, news recommenders and interactive social networks \cite{Damaskinos20}, healthcare applications \cite{Li20}, and so on. 
Recent works in \cite{Zhao18,Nishio18,Amiri19c,Barnes20,Dhakal20, Amiri20b, Chang20} study communication-efficient FL frameworks suitable for the limited communication between the PS and the clients considering varying channel conditions, quantization and sparsification, non-i.i.d.~client datasets, and coding.

The performance of different FL frameworks are usually determined by their convergence performance, average iteration time, and number of iterations. In certain FL applications such as social media networks and human mobility prediction where large quantities of highly temporal data are produced which diminish in value in a matter of hours, timeliness is also critical to incorporate rapidly changing data into the model in a timely manner. To illustrate this, we consider two clients, Alice and Bob, who participate in a next place forecasting task that aims to jointly train clients to predict their next location based on their current location and past habits. Such information is used to offer an enhanced experience and better recommendations in location-based social networks such as Twitter and Yelp as well as improved navigation that uses less congested routes. Assuming Bob gets moving earlier than Alice, to predict Alice's movements more accurately and consequently deliver the most relevant suggestions to her, Bob's (and all other similar clients') earlier activity should be incorporated into the model by the time Alice gets moving.

Motivated by this, in this work, we use the age of information metric to characterize information freshness in an FL framework. Introduced in \cite{Kaul12a}, age of information has been studied in the context of queueing networks, scheduling and optimization, energy harvesting, and so on (see the survey in \cite{SunSurvey}). Recently, age of information has found new applications in reinforcement learning and distributed computation and learning \cite{Elmagid18, Liu18, Ceran18, Beytur19, Elmagid19, Buyukates19c, Ozfatura20a, Yang19a}. Particularly, \cite{Yang19a} studies an age metric, called age of update, in the FL context to measure the staleness of each update and schedule clients at each iteration accordingly. Based on this age-based metric, authors in \cite{Yang19a} propose a scheduling policy, which takes the staleness and communication quality of devices into account to accelerate the convergence of FL. In \cite{Yang19a}, age is used as a client scheduling metric rather than an end-to-end performance metric. 

In this work, we propose a timely communication framework for FL that considers limited client availability and communication resources to make sure that locally generated client data are incorporated in the learning model with as little age as possible. In the proposed scheme, the PS waits for $m$ available clients out of the total $n$ clients at each iteration and uses local updates of the earliest $k$ clients among these $m$ available ones to update the global model. When $k$ is larger, more clients update the PS at the expense of larger iteration durations. To obtain a larger $k$ value, we can increase $m$ which induces larger waiting times for client availability. In such a system, we characterize the average age experienced by each client and determine the age-optimal $k$ and $m$ values. We show that, in addition to ensuring freshness, the proposed timely communication scheme significantly improves the average iteration time compared to random client selection employed by \cite{McMahan17} without harming the convergence of the global model. 

\section{System Model}
In a typical FL setup with a single PS and $n$ clients, each client $j$ has the local data set $\mathcal{B}_j$, with $N_j = |\mathcal{B}_j |$ (see Fig.~\ref{fig:FL_update}). The aim of the PS is to train a model parameter vector $\boldsymbol{\theta} \in \mathbb{R}^d$ using the data stored locally across the clients to minimize a particular loss function given by $L(\boldsymbol{\theta}) = \frac{1}{n} \sum_{j=1}^{n} L_j (\boldsymbol{\theta})$,
where $L_j (\boldsymbol{\theta})$ denotes the application specific loss function at client $j$ and is computed over the $N_j$ samples in $\mathcal{B}_j$, $j \in [n]$. 

At each iteration $t$, each client receives the current model $\boldsymbol{\theta}_t$ from the PS and performs $\tau$-step stochastic gradient descent (SGD) for $\tau \in \mathbb{N}$ to minimize an empirical loss function with respect to its local dataset by using $\boldsymbol{\theta}_t$. At iteration $t$, the $\ell$th step of the local SGD is given by  
\begin{equation}
\boldsymbol{\theta}^{j}_{\ell+1}=\boldsymbol{\theta}^{j}_{\ell}-\eta^j_\ell \nabla L_j(\boldsymbol{\theta}^j_{\ell}), \quad \ell\in[\tau] \label{mod_update}
\end{equation}
where $\eta^j_\ell$ is the learning rate. Each selected device sends its model estimate after $\tau$ local steps, denoted by $\boldsymbol{\theta}^j_{t+1}$, with $\boldsymbol{\theta}^j_{t+1}=\boldsymbol{\theta}^j_{\tau+1}$, 
to the PS which updates the global model using
\begin{align}
	\boldsymbol{\theta}_{t+1} = \sum_{j=1}^{n} \frac{N_j}{N} \boldsymbol{\theta}^j_{t+1}. \label{FL_update}
\end{align}
Then, the updated global model $\boldsymbol{\theta}_{t+1}$ is shared with the clients and the whole process is repeated until convergence. 

In our model, clients are not always available for the learning task.\footnote{This is because, to participate in a learning task that potentially includes heavy computations, devices need to be plugged-in and running on an unmetered WiFi connection.} Rather, clients experience exponentially separated \emph{availability windows} to participate in the learning task. We assume that once available each client completes an iteration after which the next availability window of that client starts in an exponential duration of time. At each iteration, the PS schedules $m$ clients such that upon completion of an iteration, the PS waits for $Z_{m:n}$ duration of time\footnote{ We denote the $m$th smallest of random variables $Z_1,\ldots,Z_n$ as $Z_{m:n}$.} to have $m$ available clients for the next iteration, where $Z$ is an exponential random variable with rate $\lambda$ from the memoryless property.\footnote{To model the case in which the clients are all available, we can take $\lambda \rightarrow \infty$ in which case the PS selects $m$ clients randomly at each iteration.}

The set $\mathcal{A}^m_t$ denotes the first $m$ available clients in iteration $t$ to which the PS broadcasts the current model, $\boldsymbol{\theta}_t$.\footnote{Clients in $\mathcal{A}^m_t$ commit to participate in iteration $t$ such that a client in $\mathcal{A}^m_t$ does not become unavailable within the $Z_{m:n}$ duration while waiting for others to become available.} Link delays in downlink transmissions from the PS to the clients are modeled with an exponential random variable with rate $\mu$. 
We assume that the actual computation duration at clients is a deterministic duration $c$ so that the time in between the beginning of an iteration until a participating client generates its local update is a shifted exponential random variable $X$ with rate $(c, \mu)$.\footnote{We note that each selected available client performs a local SGD using the same number of samples at each local step. When clients have identical computation capabilities we have the same computation time $c$ for each client.} The link delay back to the PS in uplink transmissions is an exponential random variable $\tilde{X}$ with rate $\tilde{\mu}$. Once the PS collects the earliest $k$ of the $m$ local updates, $k\leq m$, it updates the global model as in (\ref{FL_update}). We denote the set of the earliest $k$ clients out of the available $m$ clients at iteration $t$ with $\mathcal{A}^k_t$ such that $|\mathcal{A}^k_t| = k$ and $\mathcal{A}^k_t \subseteq \mathcal{A}^m_t$.

Our aim is to design a learning framework, where the locally generated data are incorporated at the PS as timely as possible. In this work, we consider the age of information metric to measure the timeliness of the received information. The PS keeps the age of information of each client. A client's age is updated whenever its local update is received by the PS. Otherwise, its age continues to increase linearly in time. We define the long term average age of a client $j$ as
\begin{align}
   \Delta_j = \lim_{T\to\infty} \frac{1}{T}\int_0^T\Delta_j(\Bar{t})d\Bar{t},
\end{align}
where $\Delta_j(\Bar{t})$ represents the instantaneous age of client $j$ at time $\Bar{t}$ at the PS. We have $\Delta_j(\Bar{t}) = \Bar{t} - u_j(\Bar{t})$, where $u_j(\Bar{t})$ denotes the generation time of the most recent local update of client $j$ that is available at the PS. We note that from the i.i.d.~nature of this system, each client experiences identical age of information. Thus, in what follows, we focus on a single client and calculate its average age of information.

\section{Average Age Analysis}\label{sect:age}
The total time of an iteration, denoted by $Y$, is given by
\begin{align}
	Y = S + Z_{m:n}, 
\end{align}
where $S$ denotes the service time and is given by $S = (X+\tilde{X})_{k:m}$. A client participates in an iteration with probability $p_1 = \frac{m}{n}$ since the PS only selects the earliest available $m$ clients. Out of these $m$ clients only the earliest $k$ of them actually send back their updates to the PS for aggregation at each iteration. That is, given selected for the iteration, the probability of updating the PS is $p_2 = \frac{k}{m}$. Thus, at each iteration a client updates the PS with probability $p \triangleq p_1p_2 = \frac{k}{n}$. The number of iterations in between two consecutive updates from a particular client is given by a geometric random variable $M$ with rate $p$.

\begin{figure}[t]
	\centering  \includegraphics[width=0.9\columnwidth]{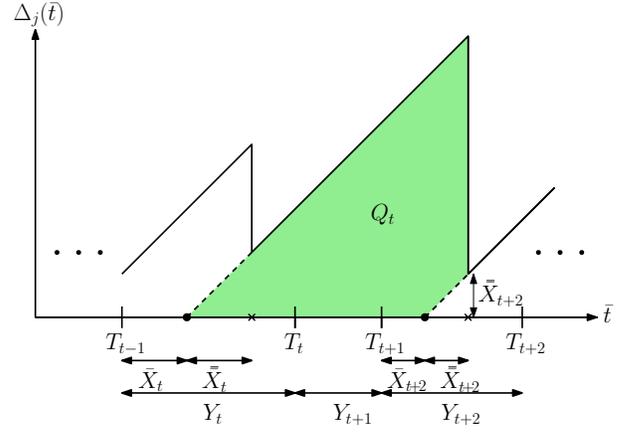}
	\caption{Sample age evolution at client $j$. Iteration $t$ starts at time $T_{t-1}$. Filled circles and crosses show the time instances at which client $j$ generates its local update and the PS receives that local update, respectively. Here, client $j$ successfully delivers its local update to the PS in iterations $t$ and $t+2$. }
	\label{fig:age_evol}
	\vspace{-0.5cm}
\end{figure}

A sample age evolution of client $j$, $j \in [n]$ is shown in Fig.~\ref{fig:age_evol}. Here, iteration $t$ starts at time $T_{t-1}$ when the PS starts broadcasting the $t$th model update to the selected available clients. Filled circles and crosses in Fig.~\ref{fig:age_evol} show the time instances at which client $j$ generates its local updates and the PS receives those corresponding local updates, respectively. We note that upon delivery, the age of client $j$ drops to the age of the most recent local update from client $j$.

In the ensuing analysis, moments of $S$ are not easy to calculate as $S$ is equal to an order statistic of a sum of exponential and shifted exponential random variables. To simplify, we assume that downlink transmissions are instantaneous since, in general, connection speeds are significantly asymmetric such that downlink transmissions are much faster than uplink transmissions \cite{Konecny16}. In this case, the time it takes for each client in $\mathcal{A}^k_t$ to generate its update is $c$ and we have $S= c + \tilde{X}_{k:m}$.

Let $\Bar{Y}$ denote the time in between the generation time of two consecutive updates from client $j$. In Fig.~\ref{fig:age_evol}, since client $j$ updates the PS in iterations $t$ and $t+2$, we have $M=2$. From Fig.~\ref{fig:age_evol}, we see that $\Bar{Y}_t = Y_t - \Bar{X}_t + Y_{t+1} + \Bar{X}_{t+2}$, where $\Bar{X}_t$ denotes the downlink delay of a client that is one of the earliest $k$ clients to deliver its update at iteration $t$, i.e., $\Bar{X} = X_j | j \in \mathcal{A}^k_t$. Since $\Bar{X}_{t} = \Bar{X}_{t+2}=c$, we have $\Bar{Y} = Y_1 + Y_2$ for $M=2$.
Then, in general, we have
\begin{align}
    \Bar{Y} = \sum_{i=1}^{M} Y_{i},\label{Y_bar}
\end{align}
which is equal to the length of the shaded trapezoid in Fig.~\ref{fig:age_evol}.

The metric we use, long term average age, is the average area under the age curve which is given by \cite{Najm17}
\begin{align}
	\Delta = \limsup_{T\to\infty} \frac{\frac{1}{T}\sum_{t=1}^{T} Q_t}{\frac{1}{T}\sum_{t=1}^{T}\Bar{Y}_t} = \frac{\mathbb{E}[Q]}{\mathbb{E}[\Bar{Y}]}. \label{avg_age1}
\end{align}
By using Fig.~\ref{fig:age_evol}, we find $ Q_t = \frac{1}{2}\Bar{Y}^2_t + \Bar{Y}_t \Bar{\Bar{X}}_{t+2}$. Thus, (\ref{avg_age1}) is equivalent to
\begin{align}
	\Delta_j = \mathbb{E}[\Bar{\Bar{X}}] + \frac{\mathbb{E}[\Bar{Y}^2]}{2\mathbb{E}[\Bar{Y}]},\label{avg_age2}
\end{align}
where $\Bar{\Bar{X}}$ denotes the uplink delay of a client that is one of the earliest $k$ clients to deliver its update at iteration $t$, i.e., $\Bar{\Bar{X}} = \tilde{X}_j | j \in \mathcal{A}^k_t$. From (\ref{Y_bar}), we find the first and second moments of $\Bar{Y}$ in terms of $Y$ as
\begin{align}
    \mathbb{E}[\Bar{Y}] &=\mathbb{E}[M]\mathbb{E}[Y] \label{EY}\\
    \mathbb{E}[\Bar{Y}^2] &= \mathbb{E}[M]\mathbb{E}[Y^2] + \mathbb{E}[Y]^2\mathbb{E}[M^2-M]. \label{EY2}
\end{align}

Inserting (\ref{EY}) and (\ref{EY2}) in (\ref{avg_age2}), we find
\begin{align}
    \Delta_j = \mathbb{E}[\Bar{\Bar{X}}] + \frac{\mathbb{E}[M^2]}{2\mathbb{E}[M]}\mathbb{E}[Y] + \frac{Var[Y]}{2\mathbb{E}[Y]}.\label{avg_age3}
\end{align}

Theorem~\ref{thm1} characterizes the average age of a client under the proposed timely communication scheme using (\ref{avg_age3}).

\begin{theorem}\label{thm1}
 Under the proposed timely communication scheme, the average age of a client is
  \begin{align}
     \Delta_j =& \frac{1}{k} \sum_{i=1}^{k} \mathbb{E}[\tilde{X}_{i:m}]  + \frac{2n-k}{2k}(c+\mathbb{E}[\tilde{X}_{k:m}] + \mathbb{E}[Z_{m:n}]) \nonumber\\ &+ \frac{ Var[\tilde{X}_{k:m}] + Var[Z_{m:n}]}{2(c+\mathbb{E}[\tilde{X}_{k:m}] + \mathbb{E}[Z_{m:n}])}.\label{avg_age5}
 \end{align}
\end{theorem}
\begin{Proof}
     We substitute the first two moments of $M$ into (\ref{avg_age3}) and note that random variables $S$ and $Z_{m:n}$ are mutually independent to obtain
\begin{align}
     \Delta_j =& \mathbb{E}[\Bar{\Bar{X}}] + \frac{2n-k}{2k}(\mathbb{E}[S] + \mathbb{E}[Z_{m:n}]) \nonumber\\ &+ \frac{Var[S] + Var[Z_{m:n}]}{2(\mathbb{E}[S] + \mathbb{E}[Z_{m:n}])}.\label{avg_age4}
 \end{align}
	The first term in (\ref{avg_age4}) is equal to
	\begin{align}
	\mathbb{E}[\Bar{\Bar{X}}] = \mathbb{E}[\tilde{X}_j| j \in \mathcal{A}^k_t] =& \sum_{i=1}^{k} \mathbb{E}[\tilde{X}_{i:m}] Pr[j=i| j \in \mathcal{A}^k_t ] \nonumber \\ =& \frac{1}{k} \sum_{i=1}^{k} \mathbb{E}[\tilde{X}_{i:m}].\label{exp}
	\end{align}	
	Here, noting that downlink transmissions are instantaneous, (\ref{exp}) follows from the fact that the earliest $k$ out of $m$ available clients is determined in an i.i.d.~fashion at a certain iteration $t$. Together with (\ref{exp}), inserting $S = c + \tilde{X}_{k:m}$ in (\ref{avg_age4}) and noting that $Var(S) = Var(\tilde{X}_{k:m})$ yield the result.
\end{Proof}


%

Next, we determine the average age of a client when the number of clients $n$ is large. Here, we note that \cite{David03}
\begin{align}
\mathbb{E}[Z_{m:n}] =& \frac{1}{\lambda}(H_n - H_{n-m}), \label{ord1} \\
Var[Z_{m:n}] =& \frac{1}{\lambda^2}(G_{n} - G_{n-m}), \label{ord2}
\end{align}
where $H_n = \sum_{j=1}^{n} \frac{1}{j}$ and $G_{n} = \sum_{j=1}^{n} \frac{1}{j^2}$. First moment and variance of $\tilde{X}_{k:m}$ in (\ref{avg_age5}) follow from (\ref{ord1}) and (\ref{ord2}) using $\tilde{\mu}$.

\begin{corollary}\label{corr_large}
For large $n$, we set $m = \alpha n$ with $m<n$ and $k=\beta m$ with $k<m$. Then, the average age of a client given in (\ref{avg_age5}) can be approximated as 
\begin{align}
    \Delta_j \approx& \frac{1}{\tilde{\mu}} + \frac{(2-\alpha\beta)c}{2\alpha\beta}-\frac{2-\alpha\beta}{2\alpha\beta\lambda}\log(1-\alpha)\nonumber\\ &+ \frac{\alpha(2-\beta)-2}{2\alpha\beta\tilde{\mu}}\log(1-\beta).\label{age_large}
\end{align}
\end{corollary}

\begin{Proof}
    Let $\delta_1$, $\delta_2$, and $\delta_3$ denote the terms in (\ref{avg_age5}). Then,
    \begin{align}
        \delta_1 =& \frac{1}{k} \sum_{i=1}^{k} \mathbb{E}[\tilde{X}_{i:m}] = \frac{1}{\tilde{\mu}}H_m-\frac{1}{k\tilde{\mu}}\sum_{i=1}^k H_{m-i} \label{delta1}\\ =& \frac{1}{\tilde{\mu}}-\frac{m-k}{k\tilde{\mu}}(H_m-H_{m-k})\label{delta2}\\
        \approx& \frac{1}{\tilde{\mu}} + \frac{1-\beta}{\beta\tilde{\mu}}\log(1-\beta),\label{delta3}
    \end{align}
    where (\ref{delta1}) follows from the order statistics in (\ref{ord1}). To obtain (\ref{delta2}), we use the series identity $\sum_{i=1}^k H_i = (k+1)(H_{k+1}-1)$ \cite{Zhong17a, Buyukates18b}, and (\ref{delta3}) follows from the fact that for large $n$, $H_i \approx \log(i) + \gamma$, where $\gamma$ is the Euler-Mascheroni constant and is ignored here for brevity. Also,
    \begin{align}
        \delta_2 &= \frac{2n-k}{2k}(c+\mathbb{E}[\tilde{X}_{k:m}] + \mathbb{E}[Z_{m:n}]) \\ &\approx \frac{2-\alpha\beta}{2\alpha\beta} \left(c-\frac{1}{\tilde{\mu}}\log(1-\beta)-\frac{1}{\lambda}\log(1-\alpha)\right).
    \end{align}
    Next, we have 
    \begin{align}
        \delta_3 = \frac{ Var[\tilde{X}_{k:m}] + Var[Z_{m:n}]}{2(c+\mathbb{E}[\tilde{X}_{k:m}] + \mathbb{E}[Z_{m:n}])} \approx 0, \label{delta3_1}
    \end{align}
    where (\ref{delta3_1}) follows by using the fact that for large $n$, $G_n \approx \frac{\pi^2}{6}$ and hence, we have $\mathbb{E}[\tilde{X}^2_{k:m}] \approx (\mathbb{E}[\tilde{X}_{k:m}])^2$ and $\mathbb{E}[\tilde{Z}^2_{m:n}] \approx (\mathbb{E}[\tilde{Z}_{m:n}])^2$ when $m$ is linear in $n$ and $k$ is linear in $m$. Summing $\delta_1$, $\delta_2$, and $\delta_3$ yields the result.
\end{Proof}

Even if the PS updates the age of $k=\alpha \beta n$ clients at each iteration, the average age expression in (\ref{age_large}) has terms that depend on the multiplication $\alpha \beta$ as well as terms that only depend on either $\alpha$ or $\beta$. Thus, to minimize the average age, we need to optimize both $\alpha$ and $\beta$ values. When $\alpha$ increases, more clients can update the PS at each iteration at the expense of longer waits for client availability. Similarly, for a given $\alpha$, when $\beta$ increases, more clients update the PS at the expense of longer iterations. Thus, parameters of $\alpha$ and $\beta$ need to be carefully selected to obtain good age performance.

\section{Numerical Results}

In this section, we provide numerical results to determine the age-optimal $\alpha$ and $\beta$ values that minimize the average age of the clients. In our simulations, we have $n=100$ clients. In the first three simulations, we plot the average age of a client as a function of $k$ for the age-optimal $m$ value. That is, we first find the age-optimal $(m,k)$ pair (equivalently, the age-optimal $(\alpha,\beta)$ pair) and then plot the average age of a client as a function of $k$ when using the age-optimal $m$ value.


In the first simulation, we take $\lambda=1$, $c=1$, and vary $\tilde{\mu}$. We observe that the average age decreases with increasing uplink transmission rates $\tilde{\mu}$. We find that the age-optimal $m$ values are $95,94,92,90,86$ for $\tilde{\mu}=0.1, 0.2, 0.5, 1, 5$, respectively. This shows that with increasing $\tilde{\mu}$, i.e, shorter average uplink transmission delays, the PS can afford to wait for less clients to become available at the beginning of iterations such that the age-optimal $\alpha$ decreases. This is because, as the transmissions become faster, the PS obtains the client updates quickly and the initial wait for client availability becomes the performance bottleneck. The corresponding age-optimal $k$ values are $55,64,74,79,83$ such that as the uplink transmissions become faster the PS opts for waiting more results from the clients, i.e., increasing $\beta$, instead of waiting for clients to become available in the next iteration. Further, more specifically for the low transmission rates, i.e., $\tilde{\mu}=0.1, 0.2, 0.5$ cases, we have a familiar age curve as in the earlier works on multicast networks \cite{Zhong17a, Zhong18b, Buyukates18, Buyukates18b, Buyukates19} that employ an earliest $k$ out of $m$ idea to facilitate timely updates. In particular, we see that the average age first decreases when $k$ increases. This is because with increasing $k$, clients update the PS more frequently. When $k$ increases beyond a certain value, however, the average age starts to increase indicating that the PS waits for clients with slower links to perform the iteration. 

\begin{figure}[t]
	\centering  \includegraphics[width=0.72\columnwidth]{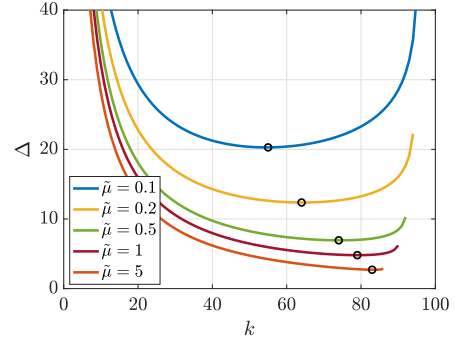}
	\caption{Average age experienced by a client as a function of $k$ with $n=100$, $\lambda=1$, and $c=1$ for varying $\tilde{\mu}$. In each curve we use the age-optimal $m$. The age-optimal $k$ values are shown with a circle. }
	\label{fig:sim1}
	\vspace{-0.5cm}
\end{figure}

\begin{figure}[t]
	\centering  \includegraphics[width=0.73\columnwidth]{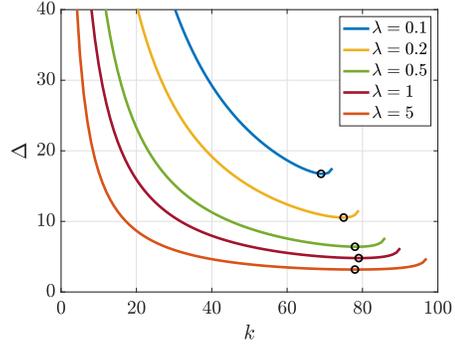}
	\caption{Average age experienced by a client as a function of $k$ with $n=100$, $\tilde{\mu}=1$, and $c=1$ for varying $\lambda$. In each curve we use the age-optimal $m$. The age-optimal $k$ values are shown with a circle. }
	\label{fig:sim2}
	\vspace{-0.5cm}
\end{figure}

In the second simulation, we consider the same setup as in Fig.~\ref{fig:sim1} but take $\tilde{\mu}=1$ and vary $\lambda$. We observe that the average age decreases for larger values of $\lambda$ as the time it takes for clients to become available is less for larger $\lambda$ values. Here, the age-optimal $m$ values are $72,79,86,90,97$ for $\lambda=0.1, 0.2, 0.5, 1, 5$, respectively, which indicate that, to facilitate timeliness, the PS selects a smaller $\alpha$ when the availability of the clients is more scarce. Corresponding age-optimal $k$ values are $69,75,78,79,78$ such that we observe that, as the clients become more frequently available, the PS uses a smaller fraction of the available $m$ clients at each iteration, i.e., $\beta$ decreases with increasing $\lambda$. In this case, instead of waiting for clients with slower links to return their update, the PS chooses to start a new iteration.

\begin{figure}[t]
	\centering  \includegraphics[width=0.73\columnwidth]{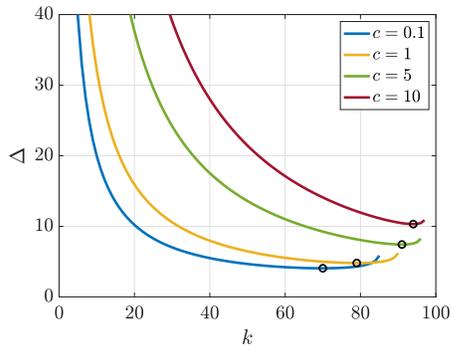}
	\caption{Average age experienced by a client as a function of $k$ with $n=100$, $\lambda=1$, and $\tilde{\mu}=1$ for varying $c$. In each curve we use the age-optimal $m$. The age-optimal $k$ values are shown with a circle. }
	\label{fig:sim3}
	\vspace{-0.5cm}
\end{figure}

In the third simulation, we consider the same setup as in Fig.~\ref{fig:sim1} but take $\tilde{\mu}=1$ and vary $c$. In this case, the age-optimal $m$ values are $85,90,96,97$ for $c=0.1, 1, 5, 10$, respectively. Corresponding age-optimal $k$ values are $70,79,91,94$. We observe from Fig.~\ref{fig:sim3} that as $c$ increases both the average age and the age-optimal $(k,m)$ values, correspondingly $(\alpha,\beta)$ values, increase. This suggests that when the fixed computation duration at the clients is larger, at each iteration, the PS is more incentivized to wait for more clients to return their updates.

\begin{figure}[t]
	\centering  \includegraphics[width=0.73\columnwidth]{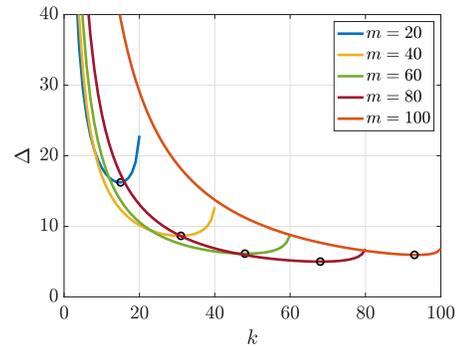}
	\caption{Average age experienced by a client as a function of $k$ with $n=100$, $\lambda=1$, $c=1$, and $\tilde{\mu}=1$ for varying $m$. The age-optimal $k$ values are shown with a circle. }
	\label{fig:sim4}
	\vspace{-0.45cm}
\end{figure}

So far, we have found the age-optimal $m$ and $k$ values to minimize the average age of the clients. In practical systems, as in \cite{Yang19a}, the PS can only schedule a fixed number of clients at each iteration due to limited communication resources such as number of subchannels etc. To investigate such settings, in the fourth simulation, we fix $m$ and analyze the average age performance by varying $k$. In Fig.~\ref{fig:sim4} we observe that the age-optimal $k$ value increases with increasing $m$. That is, when the PS waits for a larger number of available clients at each iteration, it is more beneficial for decreasing the age to wait for more of those available clients to return their updates. Here, the age-optimal $k$ values are $15,31,48,68,93$ for $m=20,40,60,80,100$, respectively. Among these $m$ values, $m=80$ gives the best average age result whereas $m=60$ and $m=100$ yield similar performance. Thus, having more communication resources is advantageous but as $m$ increases the time spent in waiting for client availability starts to hurt the age performance.

Until now, we have investigated the age-optimal $m$ and $k$ values. Our results indicate that it is not necessarily age-optimal to get updates from each client at each iteration. In particular, we show that to get more timely updates it is better to wait for the first $m$ available clients and then use the updates of the earliest $k$ clients among these $m$ available ones. 

\begin{figure}[t]
	\centering  \includegraphics[width=0.73\columnwidth]{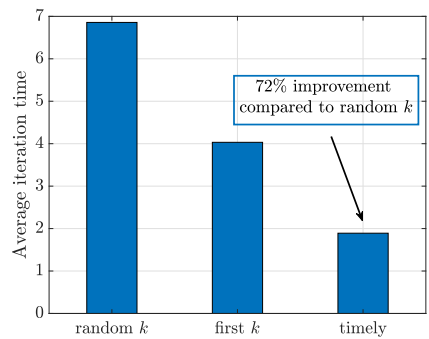}
	\caption{Average iteration time under different schemes when $k=10$, $m=20$, $n=100$ for $c=1$, $\tilde{\mu}=1$, $\lambda=1$ averaged over $50000$ iterations.}
	\label{fig:sim5}
	\vspace{-0.5cm}
\end{figure}

Next, we analyze the average iteration time $\mathbb{E}[Y]$ under the proposed timely communication framework. We compare its performance with two baseline schemes: random $k$, which selects any $k$ clients uniformly at random at each iteration and first $k$, which selects the first $k$ clients that become available at each iteration. We take $k=10$ and $m=20$ and use the same setup as in Fig.~\ref{fig:sim1}. In Fig.~\ref{fig:sim5} we see that the proposed timely communication framework outperforms the random $k$ and first $k$ schemes. In this case, the performance improvement compared to the random $k$ scheme is $72\%$. This is because random $k$ does not consider the availability of the clients to make the client selection whereas the proposed scheme uses the client availability as well as the link delays to make the client selection. We also note that even if the clients are all available, i.e., $\lambda$ tends to $\infty$, the proposed scheme still yields more than $50\%$ improvement over the random $k$ scheme. This shows that the proposed timely communication framework not only gives better age performance but also decreases the average iteration time compared to random client selection implemented in \cite{McMahan17}.

\begin{figure}[t]
	\centering  \includegraphics[width=0.73\columnwidth]{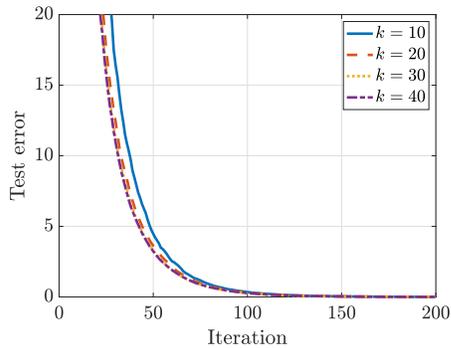}
	\caption{Convergence performance of the proposed scheme for varying $k$ with $m=40$, $n=100$ $c=1$, $\lambda=1$, and $\tilde{\mu}=1$ for a linear regression task.}
	\label{fig:conv}
	\vspace{-0.5cm}
\end{figure}


Finally, we consider the convergence performance of the proposed scheme in a learning task. The proposed timely communication operation is operationally no different than selecting a random $k$ subset of clients at each iteration uniformly at random. In other words, under the proposed operation, at each iteration, each client updates the PS with equal probability. Thus, earlier convergence results on FL that employ a random client selection at each iteration as in \cite{McMahan17} readily apply in the case of the proposed timely communication scheme. To demonstrate this, we consider a simple linear regression problem over synthetically created training and test datasets as in \cite{Ozfatura20a}. The loss function at the clients $L_j$ is the mean squared error and the size of the model is $d = 1000$. The dataset is randomly distributed to each client such that client $j$ has $N_j = 20$ samples for $j \in [n]$. We have the batch size equal to $20$, $\tau=1$, and $\eta = 0.1$ for all workers and iterations. A single simulation includes $T = 200$ iterations. Results are averaged over $5$ independent simulations.

Fig.~\ref{fig:conv} shows the convergence performance of the proposed scheme for varying $k$ and $m=40$. As shown in \cite{McMahan17} for random client selection, selecting $10\%$ of the clients, i.e., setting $k=10$, is sufficient for achieving good convergence performance. We see from Fig.~\ref{fig:sim4} that the age-optimal $k$ value is $31$ when $m=40$. Here, the age-optimal $k$ value is larger than $10$ which indicates that the proposed timely communication scheme minimizes the average age of information at the clients without slowing down the convergence.



\section{Conclusion}

In this work, we proposed a timely communication scheme for FL that is suitable for applications that include highly temporal rapidly changing client data such as social media networks, human mobility prediction systems, and news recommenders. Considering limited client availability and communication resources, in the proposed communication scheme, the PS waits until there are $m$ available clients at the beginning of each iteration. To update the global model, the PS uses the local update of the earliest $k$ clients from these available $m$ clients. Under such operation, we characterized the average age of information at the clients and numerically determined the age-optimal $m$ and $k$ values. Our results indicate that there exists an age-optimal $(m,k)$ pair that strikes a balance between waiting times for client availability, local update transmission times, and fresh model updates. We also showed that for the same $k$ value, the proposed timely communication framework significantly improves the average iteration time without hurting the convergence of the learning task.


\bibliographystyle{unsrt}
\bibliography{IEEEabrv,lib_v6}
\end{document}